\documentclass[11pt,twocolumn]{article} 
\pdfoutput=1

\usepackage[margin={1.5cm,1.5cm}]{geometry} 
\usepackage[all]{nowidow} 
\usepackage{color,soul} 

\usepackage{balance,float,tabu}  
\usepackage{subscript,gensymb,amsmath,graphicx}
\usepackage[super,sort&compress,comma]{natbib}

\newcommand{\lo}[1]{\textsubscript{#1}} 
\newcommand{\hi}[1]{\textsuperscript{#1}}

\begin{document}

\title{Understanding and Control of Bipolar Doping in Copper Nitride}

\author{Angela N.~Fioretti,\textit{$^{1,2}$} Craig P.~Schwartz,\textit{$^{3}$} John Vinson,\textit{$^{4}$} Dennis Nordlund,\textit{$^{3}$}\\ 
David Prendergast,\textit{$^{5}$} Adele C.~Tamboli,\textit{$^{1,2}$} Christopher M.~Caskey,\textit{$^{1,2}$} Filip Tuomisto,\textit{$^{6}$}\\
Florence Linez,\textit{$^{6}$} Steven T.~Christensen,\textit{$^{1}$} Eric S.~Toberer,\textit{$^{1,2}$}\\ 
Stephan Lany,\textit{$^{1}$} and Andriy Zakutayev~$^{\ast}$\textit{$^{1}$}\\
\\
\normalsize{$^{1}$National Renewable Energy Laboratory, Golden, Colorado 80401 USA}\\
\normalsize{$^{2}$Colorado School of Mines, Golden, Colorado 80401 USA}\\
\normalsize{$^{3}$Stanford Synchrotron Radiation Laboratory, SLAC National Accelerator Lab,}\\ 
\normalsize{Menlo Park, California 94720 USA}\\
\normalsize{$^{4}$National Institute of Standards and Technology, Gaithersburg, Maryland 20899 USA}\\
\normalsize{$^{5}$Lawrence Berkeley National Laboratory, Berkeley, California 94720 USA}\\
\normalsize{$^{6}$Aalto University School of Science, Espoo, Finland}\\
\\
\normalsize{$^\ast$To whom correspondence should be addressed; E-mail: andriy.zakutayev@nrel.gov.}}

\date{} 

\twocolumn[            
\begin{@twocolumnfalse}
\maketitle
\begin{abstract}
\noindent{Semiconductor materials that can be doped both n-type and p-type are desirable for diode-based applications and transistor technology. Copper nitride (Cu$_3$N) is a metastable semiconductor with a solar-relevant bandgap that has been reported to exhibit bipolar doping behavior. However, deeper understanding and better control of the mechanism behind this behavior in Cu\lo{3}N is currently lacking in the literature. In this work, we use combinatorial growth with a temperature gradient to demonstrate both conduction types of phase-pure, sputter-deposited Cu\lo{3}N thin films. Room temperature Hall effect and Seebeck effect measurements show n-type Cu\lo{3}N with 10\hi{17} electrons/cm\hi{3} for low growth temperature ($\sim$35\degree C) and p-type with 10\hi{15}--10\hi{16} holes/cm\hi{3} for elevated growth temperatures (50--120\degree C). Mobility for both types of Cu\lo{3}N was $\sim$0.1--1 cm\hi{2}/Vs. Additionally, temperature-dependent Hall effect measurements indicate that ionized defects are an important scattering mechanism in p-type films. By combining X-ray absorption spectroscopy and first-principles defect theory, we determined that V\lo{Cu} defects form preferentially in p-type Cu\lo{3}N while Cu\lo{i} defects form preferentially in n-type Cu\lo{3}N. Based on these theoretical and experimental results, we propose a kinetic defect formation mechanism for bipolar doping in Cu\lo{3}N, that is also supported by positron annihilation experiments. Overall, the results of this work highlight the importance of kinetic processes in the defect physics of metastable materials, and provide a framework that can be applied when considering the properties of such materials in general.} 
\end{abstract}
\vspace{0.5cm}
\end{@twocolumnfalse}
]

\section{Introduction}

The ability to dope a semiconductor both n-type and p-type is a critical factor in determining the viability of that material for diode-based applications and transistor technologies. In cases where bipolar doping is not achievable, selecting a suitable heterojunction partner for diode fabrication can generate a host of additional materials challenges that often hinder device development. This problem has been particularly acute in the n-type nitride-based semiconductors, such as the III-N class of materials. For example, true p-n junctions based on GaN took decades to develop, ultimately leading to a Nobel Prize.\cite{nakamura2000, maruska2015} Effective p-type doping is also problematic in solar-absorbing nitride alloys such as (In,Ga)N\cite{nakamura2000,  zhang2014} and has yet to be shown in other novel nitride semiconductors with solar-matched bandgaps, like ZnSnN\lo{2}. Given the difficulty in obtaining bipolar doping in many cases, particularly in nitride semiconductors, any material that exhibits this quality warrants pursuit of a deeper understanding of the phenomenon.

In contrast to many other nitride materials, bipolar doping has been experimentally demonstrated\cite{zakutayev2014} and externally verified\cite{matsuzaki2014} in the metastable semiconductor copper nitride (Cu$_{3}$N). This behavior has been observed as a function of either growth temperature\cite{zakutayev2014} or Cu/N flux ratio,\cite{matsuzaki2014} suggesting some kind of self-doping mechanism. The ability to achieve both conduction types in Cu\lo{3}N is particularly intriguing due to its direct bandgap of 1.4 eV and indirect bandgap of 1.0 eV, which are well-suited for photovoltaic (PV) applications.\cite{caskey2014,zakutayev2014} Growth temperature would be expected to affect the defect balance (and thus doping type) in Cu\textsubscript{3}N, since nitrogen chemical potential is a function of both nitrogen pressure and substrate temperature under the ideal gas approximation.\cite{caskey2014} However, changing Cu$_{3}$N growth temperature may not only control its doping but can also lead to its decomposition or change the surface kinetics during growth, since Cu$_{3}$N is a thermodynamically metastable material.\cite{maya1993, liu1998, caskey2014} Thus, a sound mechanistic understanding of \emph{how} bipolar doping occurs in this material is necessary in order to manipulate this property with skill, and such understanding is currently lacking in the literature. 

Prior work on Cu\textsubscript{3}N has focused on characterizing properties such as decomposition temperature,\cite{maya1993, liu1998, nosaka2001} reflectivity,\cite{asano1990, maruyama1995, cremer2000} and conductivity\cite{maruyama1995} as a function of changing deposition parameters. Cu$_{3}$N was found to exhibit either n-type conduction\cite{navio2007, cho2012} or quasi-metallic behavior,\cite{hadian2012, maruyama1995} with carrier concentration or conductivity being directly proportional to substrate temperature during growth. Theoretical calculations later showed that Cu$_{3}$N has a band structure with anti-bonding states at the valence band maximum (VBM), opposite to that of conventional semiconductors with bonding states at the VBM. Such an unusual band structure causes defects to form as shallow states near the band edges rather than in the mid-gap, leading to a property known as defect tolerance.\cite{zakutayev2014} This finding suggests that intentionally introducing native point defects into Cu$_{3}$N during growth could be a possible technique for tuning its doping without deteriorating its performance. 

In the following manuscript, we demonstrate n-type and p-type conduction in Cu$_3$N as a function of substrate temperature during growth, and propose a mechanism that explains both the origins and the temperature-dependence of this bipolar doping behavior. The findings presented in this work create a new framework for understanding bipolar doping in Cu$_{3}$N, which is strongly dependent on a balance between surface and bulk kinetic processes. This understanding is achieved using a combinatorial synthesis method paired with high throughput characterization, in addition to advanced characterization techniques and both bulk and surface defect calculations. Finally, the kinetically-driven, point-defect-based mechanism developed herein for bipolar doping in Cu\lo{3}N highlights the important influence of kinetic processes on the defect physics of metastable materials in general.

\section{Methods}

\subsection{Thin Film Deposition}

Thin films of Cu$_{3}$N were deposited via radio frequency (RF) sputter deposition on 50x50 mm Eagle XG glass substrates using a combinatorial sputter deposition chamber with 1~x~10\hi{-6}~Torr residual water base pressure and an RF-plasma atomic nitrogen source. Two sputter guns with 50 mm elemental Cu targets were inclined at 45\degree~with respect to the substrate, and the atomic nitrogen source was positioned at normal incidence angle, to achieve a nominally constant film thickness and to avoid a gradient in nitrogen activity (left side, Fig.~1). 

\begin{figure}[ht]
\centering
  \includegraphics[width=8.5cm]{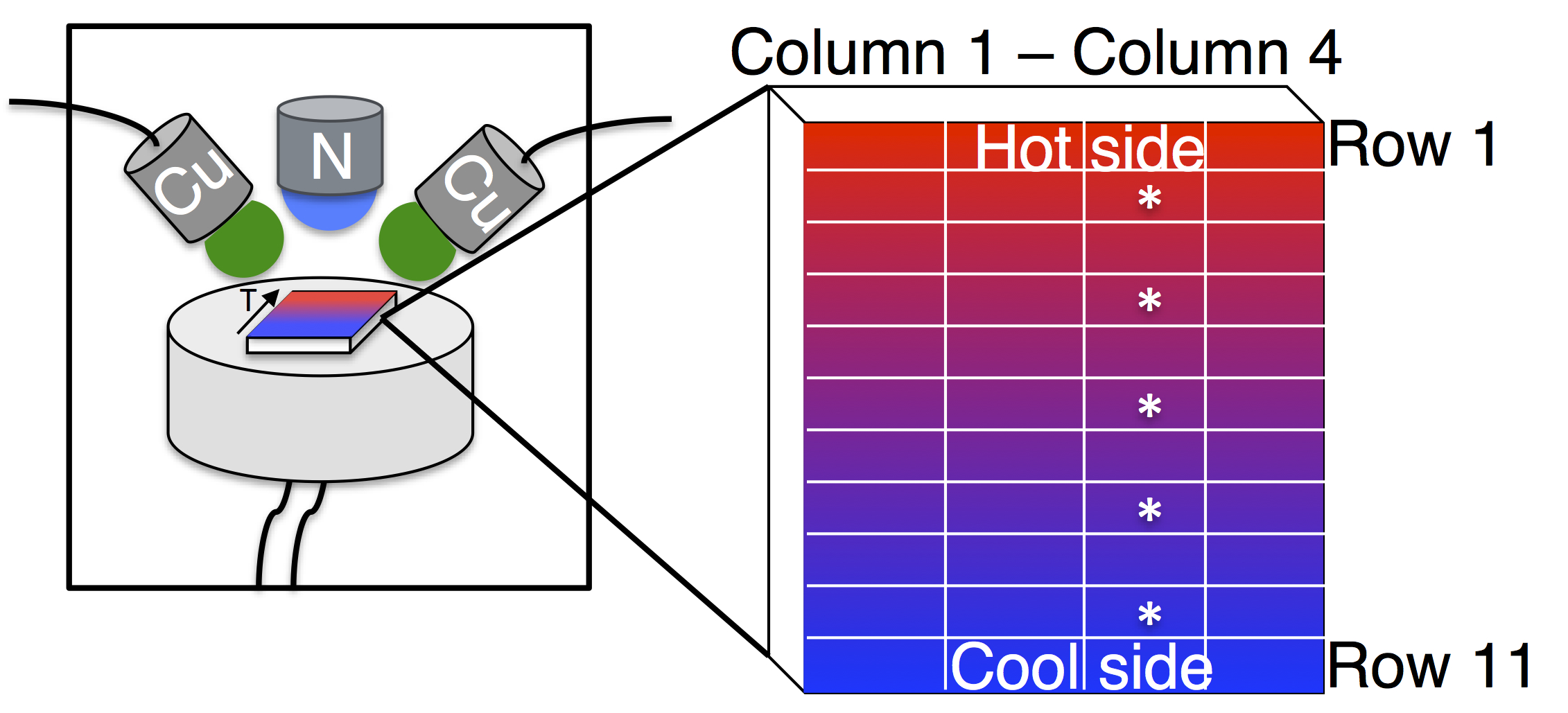}
  \caption{Diagram depicting the combinatorial chamber geometry used to obtain a gradient only in temperature (left) and the layout of the 44-point combinatorial grid used for all mapping characterization described in this work (right). The asterisks denote regions of the film that were removed and subjected to non-mapping-style measurements.}
  \label{fgr:combi}
\end{figure}

Target-substrate distance was kept at 13~cm for all samples. For a more detailed description of the atomic nitrogen source and of the sputter deposition procedures used in this work, see Ref.~[6].\nocite{caskey2014} Target power density for each gun was maintained at 1.5 W cm\textsuperscript{-2} and the atomic nitrogen source was powered at 250~W. Nitrogen gas was flowed in through the nitrogen RF source at a rate of 10~sccm and balanced by 10~sccm argon gas for a total chamber pressure of 20~mTorr. 

One side of the glass substrate was kept in contact with a resistive heating unit maintained at constant temperature, so that a continuous temperature gradient was induced across the substrate during growth. For a more detailed description of how the combinatorial temperature gradient was induced in this work, see Ref.~[16].\nocite{subramaniyan2014} Hot-side set point temperatures explored in this study were 120\degree C, 90\degree C, and no active heating. These set point temperatures represent a continuous growth temperature spread from 120--50\degree C in addition to a growth temperature of 35\degree C, as calibrated by a thermocouple on the substrate surface. 

\subsection{Characterization}

Each deposition performed as described above yielded a library of samples grown at a range of temperatures dictated by the hot-side temperature. Long-range phase purity was determined using X-ray diffraction (XRD). Conductivity of each library was measured using four-point probe sheet resistance measurements combined with thickness measured by profilometry.\cite{zakutayev2012} XRD, four point probe, and profilometry measurements were taken using a combinatorial 44-point grid (right side, Fig.~1). Film microstructure was imaged at representative regions via scanning electron microscopy (SEM). Carrier concentration and mobility were measured on selected regions of each library using room temperature Hall effect in the van der Pauw configuration and room temperature Seebeck effect. Regions selected for Hall and Seebeck measurements are indicated in Fig.~1 with white asterisks, and were removed from the original libraries in 7x7 mm pieces. Temperature-dependent Hall effect measurements were performed on a region of another p-type library grown isothermally at $\sim$160\degree C. Error bars shown for the conductivity data give the standard deviations among the four repeated rows of each combinatorial sample. Error bars shown for the Hall effect data give the standard deviations among 5--10 repeated measurements of each of the selected pieces measured from each library.  Further information on the experimental methods used herein can be found in prior publications.\cite{zakutayev2012, zakutayev2014, fioretti2015}

Experimental X-ray absorption spectra (XAS) were taken on representative p-type and n-type Cu\lo{3}N samples at beamline 8-2 at the Stanford Synchrotron Radiation Lightsource (SSRL). A nominal resolution of $\Delta$E/E of 10\hi{-4} is provided by this beamline.\cite{zhang2015} Data was collected both by using largely bulk-sensitive total fluorescence yield (TFY), which has a penetration of $\sim$100 nm, and by the more surface sensitive total electron yield (TEY), which has a penetration of $\sim$10 nm.\cite{tirsell1990} 

Additionally, both n- and p-type Cu\lo{3}N samples were analyzed using positron annihilation spectroscopy. The positron irradiation energy was tuned from 0.5 to 25 keV to probe the samples from the surface up to a mean implantation depth of 1.3 $\mu$m (i.e.~into the substrate). After implantation and thermalization in the layer, each positron diffuses and annihilates with an electron. Each such event results in two photons with energy of 511 $\pm\Delta$E keV, which are measured by a high-purity Ge detector with an energy resolution of 1.3 keV full width at half maximum. The Doppler broadening, $\Delta$E, depends on the momentum distribution of the annihilating electrons. Typically, the Doppler broadened energy spectrum narrows when positrons are trapped by vacancy defects. The shape of the spectrum is characterized using S and W shape parameters referring to the fraction of positron annihilation with low-momentum electrons (p\lo{L} \textless~0.4 a.u.) and high-momentum electrons (1.6 a.u. \textless~p\lo{L} \textless~4 a.u.), respectively.\cite{tuomisto2013}

\subsection{Computation}

The defect formation energies $\Delta H_D$ for the copper vacancy (V\lo{Cu}), copper interstitial (Cu\lo{i}), the N vacancy (V\lo{N}), and substitutional oxygen (O\lo{N}) were determined using the results of previously published supercell calculations,\cite{peng2013} using the GGA+U functional with U = 5 eV for Cu-d. In order to determine the $\Delta H_D$ for the actual growth conditions, we assumed here a quasi-equilibrium of Cu\lo{3}N with Cu-metal ($\Delta \mu_{Cu}$ = 0), and an activated source of N ($\Delta \mu_{Cu}$ = +0.76 eV). For the chemical potential of oxygen, we assumed $\Delta \mu_{Cu}$ = -0.84 eV, which corresponds to the ideal gas law chemical potential for T = 200\degree~C (upper T limit for Cu\lo{3}N growth) and pO\lo{2}=10\hi{-8} atm (base pressure of the growth chamber). In order to calculate the dependence of the V\lo{Cu} and Cu\lo{i} defect formation energy as a function of the distance from the surface, we used a slab supercell with 248 atoms modeling the partially oxidized (001) surface of Cu\lo{3}N. This surface termination, formed by substituting every other N atom of the (001) surface by an O atom was previously found to yield an exceedingly low surface energy.\cite{zakutayev2014} The charge compensation was maintained by changing the O/N ratio at the surface upon introducing the defects. These supercell calculations were performed using the VASP code,\cite{kresse1999} with otherwise identical computational settings as in Ref.~[24].\nocite{peng2013}

The modeled XAS structures were generated as detailed previously, with the exception of the CuO surface structure.\cite{zakutayev2014} The previously-calculated bulk structures consist of approximately 100 atoms with specific defects inserted into them. The surface structures consisted of at least 50 atoms. The CuO structure was generated by placing CuO on top of a slab of Cu\lo{3}N and relaxing the structure within VASP.\cite{kresse1993}  

The X-ray calculations were performed in two ways, both largely detailed previously.\cite{vinson2011,england2011} First, X-ray core hole (XCH) spectra were calculated only at the nitrogen K-edge. Within an ultrasoft pseudopotential formalism, the excitation is simulated by replacing the excited atom with a 1s core hole, such that the excited nitrogen now is represented as 1s\hi{1}2s\hi{2}2p\hi{4}. K-point sampling is used to converge the spectra. The unoccupied p-type orbitals are then projected onto the excited atom and the resultant stick intensity as a function of energy are broadened by 0.05 eV and the process is repeated for every atom.  

Second, calculations based on solving the Bethe-Salpeter equation (BSE) were performed within \textsc{ocean} for Cu L-edge as previously detailed, using DFT as a basis.\cite{vinson2014} The electronic structure was calculated within the local-density approximation using the Quantum{\sc espresso} code,\cite{espresso} and sampling of the Brillouin zone was improved through use of the OBF interpolation scheme. The number of bands included hundreds (\textgreater1000) of the first conduction bands, and the screening of the core-hole potential included thousands (\textgreater4000) bands. The k-space and mesh sampling were checked for convergence. The broadening was 0.2 eV. The spectra were aligned based on the Fermi level and potential on the core.

\begin{figure}[t!]
\hspace{-0.7em}\includegraphics[width=9.5cm]{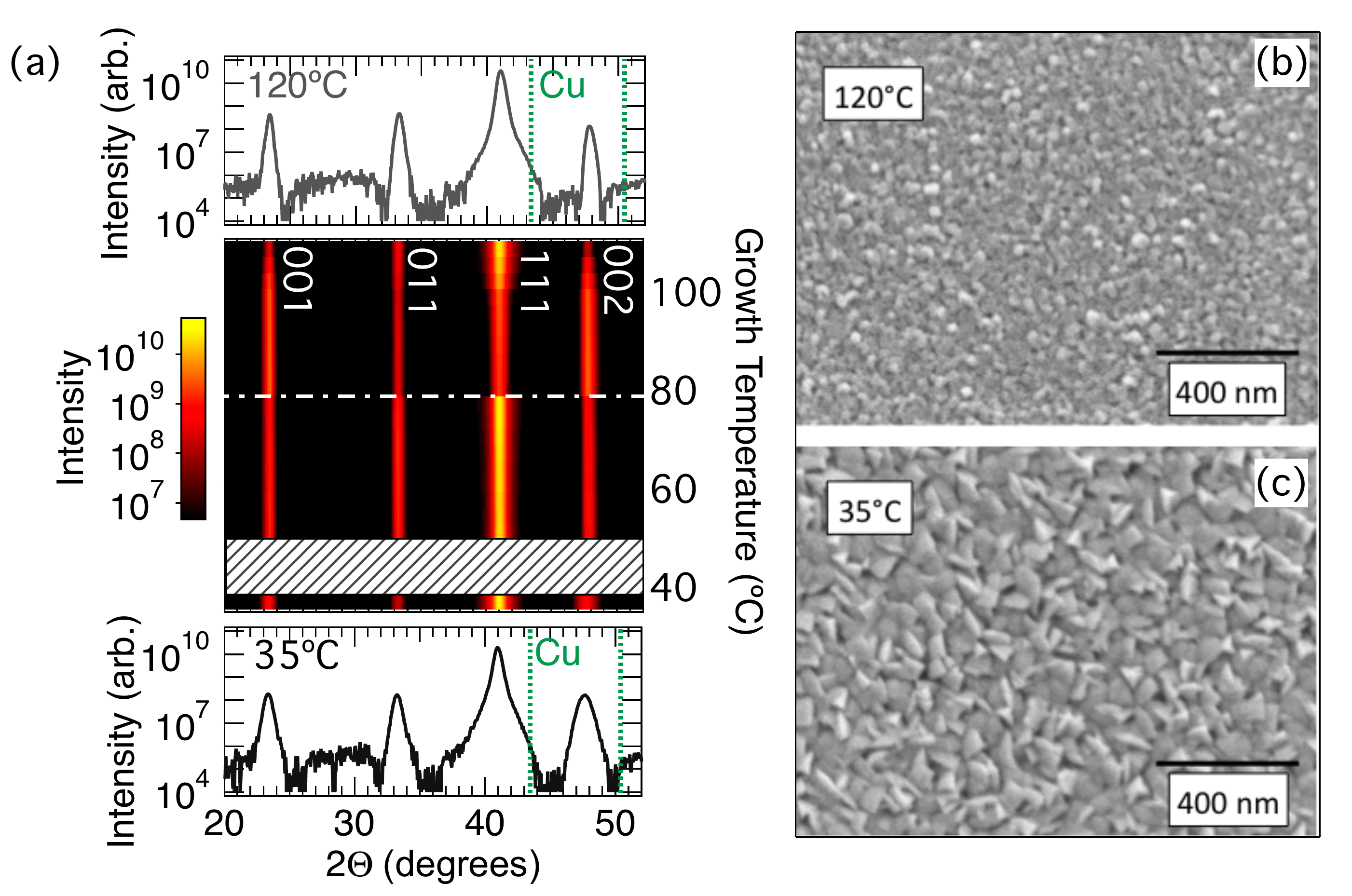}
\caption{(a) Cu$_{3}$N XRD patterns as a function of growth temperature spanning the full temperature gradient. All peaks shown correspond to Cu$_{3}$N, and no contribution from copper metal can be seen. The dashed white line horizontally separates the 120\degree C data set from the 90\degree C data set. (b-c) SEM images of characteristic Cu$_{3}$N samples grown on glass at (b) 120\degree C and (c) no active heating. The film grown with no active heating exhibits pyramidal grain termination with 50--100 nm grains while the film grown at 120\degree C exhibits spheroid grain termination with 10--50 nm grains. }
\label{fgr:xrd}
\end{figure} 

\section{Results}

\subsection{Structure}

Crystalline Cu\lo{3}N films with long-range phase-purity were achieved at all growth temperatures explored in this work. Long-range phase-purity can be seen in Fig. 2a, which displays XRD patterns for Cu$_{3}$N grown at various temperatures. Based on comparison to the peak intensities of the Cu$_{3}$N powder diffraction pattern from the Inorganic Crystal Structure Database (ICSD), these films exhibit no preferential orientation. The absence of peak intensity at 50.5\degree~and at 43.3\degree~confirms the absence of any long-range metallic copper second phase, indicating the films in this work represent a comparison between phase-pure Cu$_{3}$N films grown at different conditions and not between Cu$_{3}$N and a phase-separated mixture of Cu$_{3}$N and Cu\textsuperscript{0}. 
 
As shown in Figs.~2b-c, two distinct film morphologies were observed for the thin films prepared in this study. Films grown without actively heating the substrate (i.e. the only source of substrate heating was from the impinging sputtered atom flux), exhibited pyramidal grain termination with 50-100 nm grains (Fig.~2b). For films grown at slightly elevated substrate temperature (i.e. 50--120\degree C) spheroid grain termination was observed with grain size on the order of 10--50 nm (Fig.~2c). Observing larger grains for unheated growth and smaller grains for slightly elevated growth temperature is related to the metastable nature of Cu${_3}$N. At elevated growth temperature, slightly below the decomposition temperature of Cu\lo{3}N, it is possible that partial evaporation leads to spherical grain termination such that surface area is minimized. A more detailed description of this counterintuitive grain size phenomenon and its relation to metastability can be found in previous works.\cite{hadian2012, caskey2014}

\begin{figure}[h!]
\centerline{\includegraphics[width=8cm]{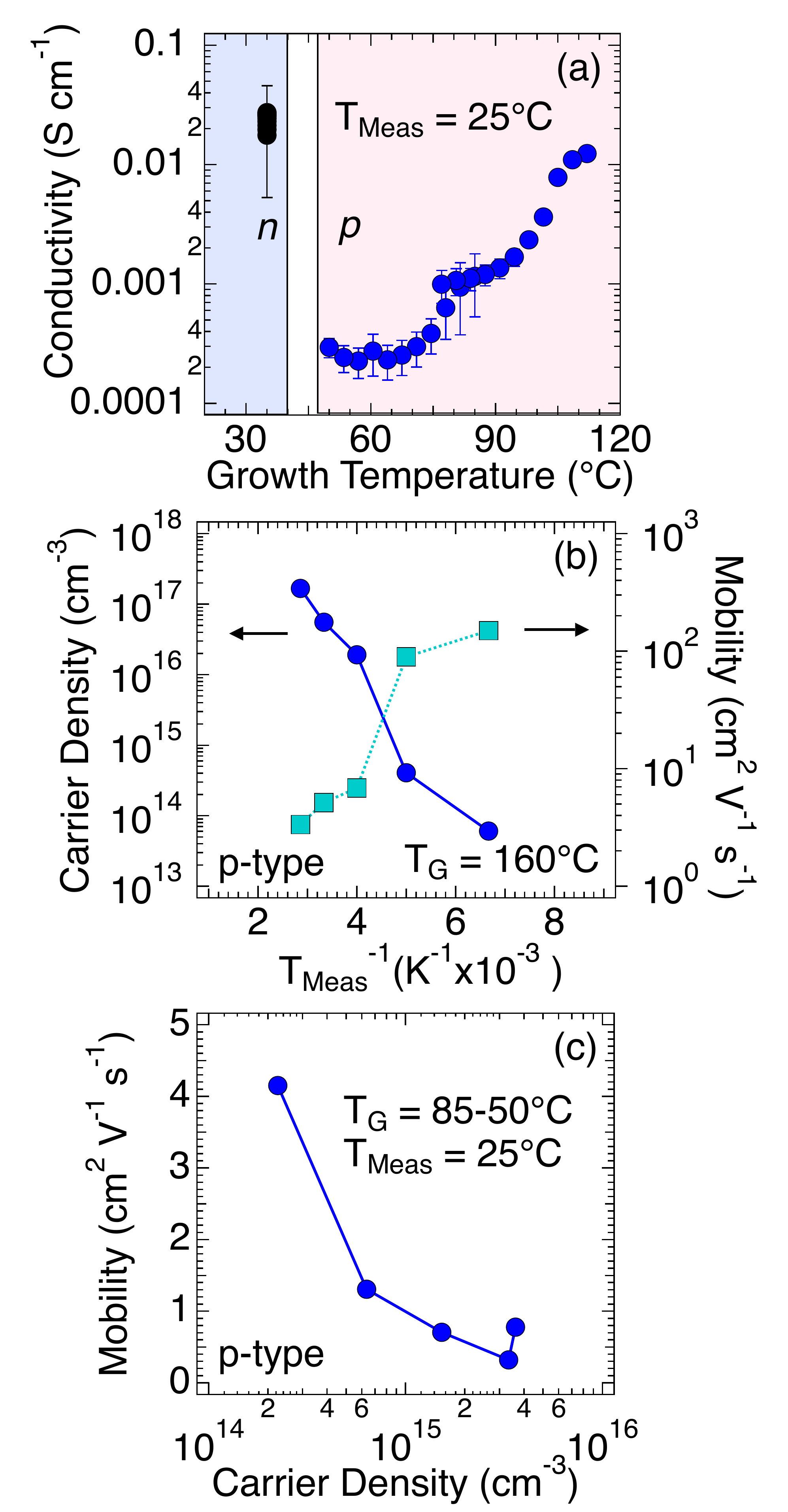}}
\caption{Cu$_{3}$N conductivity determined by four point probe as a function of growth temperature for both low and elevated growth temperature Cu\lo{3}N films is shown in (a). T-dependent Hall effect data is shown in (b) for the 160\degree film. An inverse relationship between carrier density and mobility as temperature decreases indicates ionized defect scattering dominates the room temperature mobility. For low temperature p-type films, room temperature mobility was found to decrease exponentially with increasing carrier density, as shown in (c), again indicating the influence of ionized defect scattering for p-type films. Designations of n-type and p-type in all panels are based on the sign of the Hall coefficients and corresponding Seebeck coefficients for the samples in question, summarized in Table 1.}
  \label{fgr:trans}
\end{figure}

\subsection{Transport}

Fig.~3 shows the electrical transport properties of Cu\textsubscript{3}N films as a function of growth temperature. Designations of n-type and p-type in all panels of Fig.~3 are based on the sign of the Hall coefficients and corresponding Seebeck coefficients for the samples in question. The average n-type and p-type carrier density, mobility, and Seebeck values are summarized in Table 1. Films from the 120\degree C region were unable to be reliably measured by Hall effect, due to contact degradation with repeated measurement. The same effect was not observed while collecting Seebeck effect data, which allowed the high-temperature samples to be designated p-type. To fill the gap in the Hall effect data, a sample grown isothermally at 160\degree C in a previous work\cite{zakutayev2014} was used for obtaining temperature-dependent Hall effect data. The data from that sample is indicated with an asterisk in Table~1.

\begin{table*}[t]
\centering
\begin{tabular}{||c c c c c||} 
 \hline
 Growth T / \degree C & Carrier Type & Carrier Density / cm\hi{-3} & $\mu$ / cm\hi{2} V\hi{-1} s\hi{-1} & S / $\mu$V K\hi{-1} \\ [0.7ex] 
 \hline\hline
 35 & electrons & 1.62 x 10\hi{17} &  1.63 & -69 \\ 
 50--85 & holes & 1.87 x 10\hi{15} & 1.45 & +575 \\
 90--120 & holes & -- & -- & +421 \\
 160* & holes & 5.56 x 10\hi{16} & 0.10 & +200 \\[1ex]
 \hline
\end{tabular}
\caption{Summarized Hall and Seebeck effect data from phase-pure Cu\lo{3}N films deposited at increasing substrate temperatures. Carrier type was determined based on the signs of the Hall and Seebeck coefficients. \emph{S} represents the Seebeck coefficient for each sample and $\mu$ represents the mobility determined by Hall effect for each sample. \emph{*Data taken from Ref.}~[4]}
\label{tbl}
\end{table*}

Hall and Seebeck effect measurements of the three Cu\textsubscript{3}N libraries reveal a switch from n-type to p-type conduction as growth temperature increased from 35--160\degree C. This switch in majority carrier type coincides with a U-shaped conductivity trend measured as a function of increasing growth temperature in the 35--120\degree C range (Fig. 3a). This correlation suggests competing formation of donor and acceptor defects, with donor defects dominating during unheated growth and acceptor defects dominating at elevated temperature. Consistent with this hypothesis, evidence for ionized defect scattering was observed via temperature-dependent Hall effect for p-type samples. This can be seen in Figs.~3b-c, in which carrier density and mobility were inversely proportional to each other as a function of decreasing temperature for p-type Cu\lo{3}N grown at 160\degree C (Fig.~3b). For low temperature p-type samples, it was found that mobility exhibited an exponential decrease as a function of increasing carrier density; again, indicative of ionized defect scattering and consistent with a competing defect formation hypothesis. Collectively, these findings shed light on the underlying transport properties of Cu$_3$N and also provide a reliable method for controlled bipolar self-doping in this material. 

Demonstration in this work of bipolar doping in Cu\lo{3}N confirms the reproducibility of this phenomenon as reported elsewhere,\cite{zakutayev2014,matsuzaki2014} but does not offer insight into the specific origins of this property nor to the identity of the defects giving rise to each majority carrier type. Arguably, a deeper understanding of bipolar doping in copper nitride will be necessary if this desirable property is to be manipulated with precision. Recognizing this necessity, we used different computational methods of defect identification coupled with X-ray absorption spectroscopy and positron annihilation spectroscopy to investigate the atomic-scale origins of bipolar self-doping in Cu\lo{3}N.

\begin{figure}[h!]
\centering
\includegraphics[height=6cm]{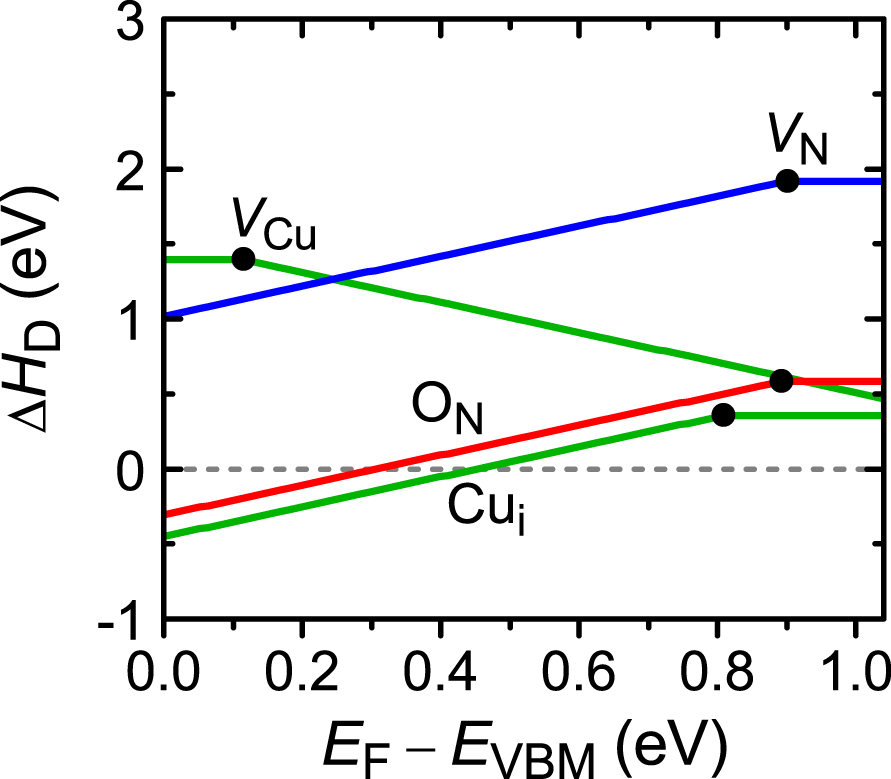}
\caption{Calculated defect formation enthalpies as a function of the Fermi level for Cu\lo{3}N, assuming a quasi-equilibrium between Cu\lo{3}N, Cu-metal, and activated N. V\lo{Cu} defects were found to be the only native point defects that could lead to p-type doping in Cu\lo{3}N. V\lo{N} defects were found to be the highest-energy defect to form among the native donor defects, indicating it is an unlikely contributor to n-type doping in this material. O\lo{N} defects were included in the calculation due to oxygen being a common impurity in nitride films.}
  \label{fgr:enth}
\end{figure}

\subsection{Bulk Defect Formation Enthalpy}

Fig. \ref{fgr:enth} shows the defect formation enthalpies as a function of the Fermi level, assuming a quasi-equilibrium between Cu\lo{3}N, Cu-metal, and activated N. Only defects that are stable and could contribute to n-type or p-type doping are shown. It is apparent from Fig.~\ref{fgr:enth} that only V\textsubscript{Cu} defects could contribute to p-type doping in this material, as this is the only defect with a positive slope as the Fermi level moves down in energy from the conduction band minimum (CBM) to the valence band maximum (VBM). This is consistent with other Cu\textsuperscript{I+}-based semiconductors, such as CuInGaS\lo{2} and Cu${_2}$O, that exhibit p-type conductivity. For n-type doping, three stable donor defects were found: V\textsubscript{N}, O\textsubscript{N}, and Cu\textsubscript{i}. Oxygen substitutional defects were included in these calculations, even though O\textsubscript{N} is not a native defect, because oxygen incorporation from background pressure is common in nitride films. The formation enthalpy of V\textsubscript{N} was found to be ~1.5 eV higher than O\textsubscript{N} and Cu\textsubscript{i} for the entire range of Fermi energies plotted, making it the least favorable donor defect to form and therefore an unlikely contributor to n-type doping in copper nitride. Considering defect formation enthalpies alone, the results presented in Fig.~\ref{fgr:enth} suggest that V\lo{Cu} defects give rise to p-type doping and O\lo{N} plus Cu\textsubscript{i} defects give rise to n-type doping.

The formation enthalpies shown in Fig.~\ref{fgr:enth} suggest that even when the incorporation of O is suppressed by reducing or eliminating O from the growth process (effectively increasing the $\Delta H_D$ of O\lo{N}), a true quasi-equilibrium situation would lead to n-type doping due to an excess of Cu\lo{i} over V\lo{Cu}. However, such a situation might be difficult to achieve, because it requires a sufficiently high temperature to enable bulk diffusion, which at the same time also facilitates the decomposition of the metastable Cu\lo{3}N compound. Therefore, we employed additional computational and experimental methods to gather more information about the atomic-scale origins of bipolar doping.

\subsection{X-Ray Absorption Spectroscopy}

\begin{figure}[t!]
\centerline{\includegraphics[width=10cm]{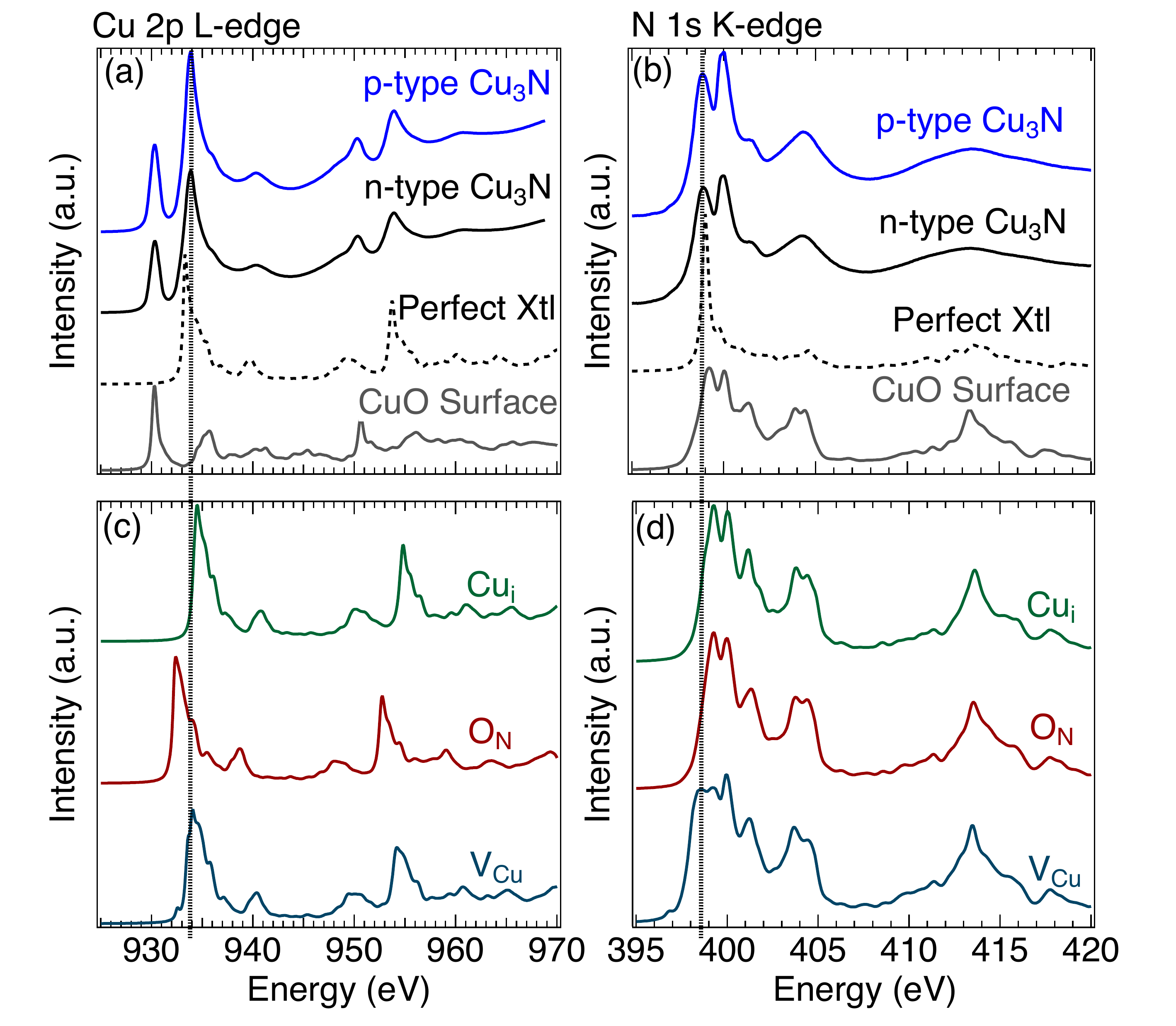}}
\caption{NEXAFS spectra for n-type (black curves) and p-type (blue curves) Cu\lo{3}N at the Cu L-edge (a) and N K-edge (b) show very similar features other than higher intensity for p-type spectra. Modeling performed using a BSE-based method shows that the experimental spectra exhibit features of a perfect Cu\lo{3}N crystal (dashed black curves) and of a Cu\hi{II+} surface state consistent with a CuO surface layer (gray curves). Modeled defect structures for Cu\lo{3}N (panels (c) and (d)) indicate that O\lo{N} substitutional defects (red curves) likely are not major contributors to the experimental NEXAFS signal, due to the shift to lower energy observed for the Cu L-edge O\lo{N} curve.}
  \label{fgr:nex}
\end{figure}

Using Near Edge X-ray Absorption Fine Structure (NEXAFS) measurements and modeling performed using {\sc ocean},\cite{vinson2011,gilmore2015} it was found that O\lo{N} defects are unlikely to be dominant contributors to bipolar doping in Cu$_3$N. This finding is in addition to the finding in Section 3.3 that V\lo{N} defects are the most energetically unfavorable donor defects to form in this material. Taken together, these results suggest that Cu\lo{i} and V\lo{Cu} defects make the most significant contributions to Cu\lo{3}N bipolar doping behavior. NEXAFS spectra for the Cu L-edge and the N K-edge for representative n-type and p-type samples are shown in Fig.~\ref{fgr:nex}, along with modeled spectra corresponding to Cu\lo{i}, V\lo{Cu}, and O\lo{N} defect structures. Both Cu\lo{i} and V\lo{Cu} defects were found to be present in both p-type and n-type Cu$_3$N, which supports the hypothesis that competing donor and acceptor defect formation is the mechanism behind bipolar self-doping in this material. 

To arrive at the above conclusions, it was necessary to use a formalism for modeling NEXAFS spectra that was not based on self-consistent field methodologies, which are often limited specifically to K-edge absorption. These methods, including FCH, X-ray core hole (XCH), and HCH, self-consistently solve the electronic structure with a core hole in place. However, the ability to model L-edge excitations using these methods is much more limited. For this reason, it was possible to model the N K-edge using XCH but not the Cu L-edge. To overcome such limitation, we instead used a method based on the Bethe-Salpeter equation (BSE), specifically \textsc{ocean},\cite{vinson2011,gilmore2015} to model both the N K-edge and the Cu L-edge NEXAFS spectra.

Upon inspection of Figs.~\ref{fgr:nex}a-b, which shows the total electron yield of the NEXAFS experiments for n- and p-type Cu\lo{3}N, it is apparent that the n-type and p-type samples produced similar spectra with only minor intensity changes. In both the Cu edge and N edge spectra, the p-type signal has greater intensity than the n-type signal. Furthermore, the modeled spectra for a perfect Cu$_3$N crystal (dashed black curves) does not fully capture the features of either the Cu L-edge or the N K-edge. Instead, the perfect crystal simulation effectively models one leading edge peak and misses the other leading edge peak in each set of spectra. It is only with the addition of a Cu\hi{II+} surface state, consistent with a CuO surface oxide (gray curves), that both leading edge peaks in both sets of spectra can be effectively modeled, indicating that a CuO surface oxide layer is likely present. It is important to note that this surface oxide layer is probably formed after growth due to exposure to air, and not formed as a native surface termination inside the growth chamber. This is in contrast to the partially oxidized (001) surface of Cu\lo{3}N described in Ref. [4],\nocite{zakutayev2014} which is anticipated to form even with minimal presence of O during growth. Both n-type and p-type Cu$_3$N exhibit peaks consistent with a CuO surface oxide layer, and also exhibit features consistent with a perfect Cu$_3$N crystal.

Three defect structures for Cu\lo{3}N were simulated using the BSE-based method described above: Cu\lo{i}, V\lo{Cu}, and O\lo{N}. These defect structures were chosen based on the point defect formation enthalpies presented in Section 3.3. The resulting modeled spectra for each edge are displayed in Figs.~\ref{fgr:nex}c-d. A vertical gray line is overlaid on each set of panels to indicate the experimental peak position of the first leading edge peak that is not due to the Cu\hi{II+} surface layer. Looking specifically at the Cu L-edge (Figs.~\ref{fgr:nex}a and c), the only defect structure that is unlikely to be a major contributor to the NEXAFS signal is O\lo{N} substitution (red curve), due to the modeled spectrum being significantly red-shifted compared to experiment. This is not to say that O\lo{N} defects are not present in the films in this work, only that O\lo{N} substitutions in the bulk must exist at a far lower concentration than either Cu\lo{i} or V\lo{Cu}. Turning to the N K-edge (Figs.~\ref{fgr:nex}b and d), it is clear that none of the other defect structures are shifted enough to be reasonably ruled out as contributing to the overall spectra.
It should be noted that the resemblance of the simulated defect spectra in panel (d) to the CuO surface layer spectrum in panel (b)  is due to the s-p transition of the nitrogen K-edge being highly sensitive to broken local symmetry.\cite{prendergast2014,uejio2008} This explains why the simulated defect spectra shown in panel (c) more closely resemble the simulated perfect crystal spectra shown in panel (a), rather than resembling the simulated CuO surface layer: because the p-d transitions probed at the Cu L-edge are not as sensitive to broken local symmetry, and therefore the defect spectra are not dominated by these effects as much as the N k-edge spectra are. The conclusion that O\lo{N} defects are not major contributors to the NEXAFS spectra is consistent with the films in this work being grown under activated nitrogen atmosphere, in which substitution of oxygen impurities on nitrogen sites is mitigated by the dramatic improvement in nitrogen incorporation.\cite{caskey2014} Overall, the combination of calculated defect formation enthalpies and modeled NEXAFS spectra points to Cu\lo{i} and V\lo{Cu} defects as being the most prominent point defects in both p-type and n-type Cu\lo{3}N.

Conceptually, if both Cu\lo{i} and V\lo{Cu} defects are present in both conduction types of Cu\lo{3}N, then it follows that the donor defect (Cu\lo{i}) must dominate in n-type copper nitride and the acceptor defect must dominate in p-type copper nitride. What is not clear from this analysis is \emph{why} the donor defect would form preferentially at low growth temperature ($\sim$35\degree C) and the acceptor defect (V\lo{Cu}) would do so at elevated growth temperature (120\degree C). To answer this question, and thus propose a complete mechanism for bipolar doping in Cu\lo{3}N, an additional set of first principles calculations were performed. Specifically, depth-dependent defect formation enthalpies were calculated to determine whether each defect type (Cu\lo{i} and V\lo{Cu}) could form preferentially at the surface or in the bulk. Based on these surface calculation results, a kinetic mechanism was proposed to explain the observed temperature dependence of the doping type in Cu\lo{3}N.

\begin{figure}[h!]
\centering
\includegraphics[width=8cm]{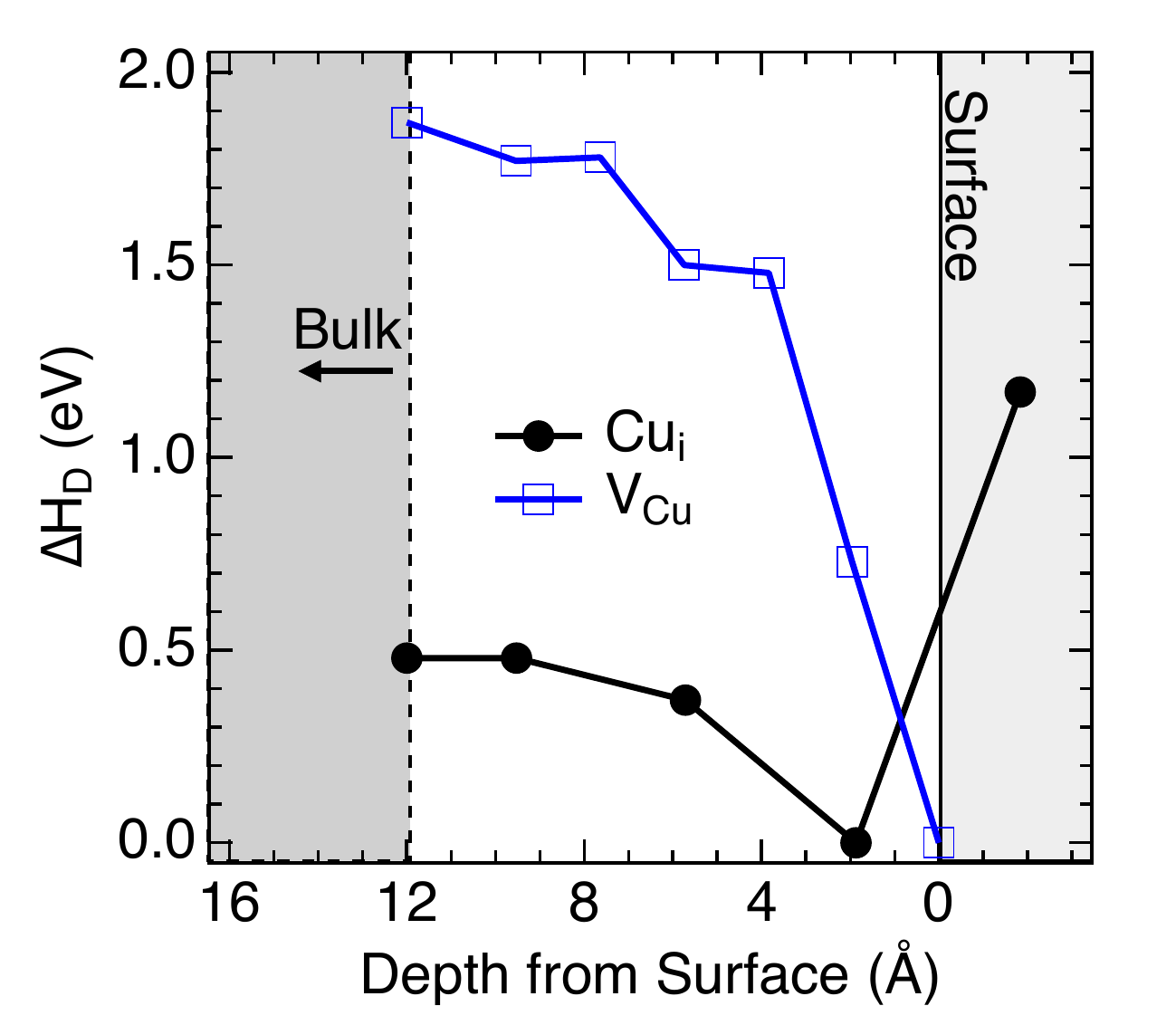}
\caption{Depth-dependent defect formation enthalpies for the donor defect, Cu\lo{i}, and for the acceptor defect, V\lo{Cu}, show that V\lo{Cu} defects form preferentially at the film surface and are high energy defects to form in the bulk. In contrast, Cu\lo{i} defects are high energy to form at the film surface but require lower energy to form in the bulk compared to V\lo{Cu}.}
  \label{fgr:dep}
\end{figure}

\subsection{Depth-dependent Defect Formation} 

To build upon the conclusions presented above, point defect formation enthalpies for Cu\lo{3}N were recalculated as a function of depth from the film surface. Only formation enthalpies for Cu\lo{i} and V\lo{Cu} defects were recalculated, because both V\lo{N} defects (Fig.~\ref{fgr:enth}) and O\lo{N} defects (Fig.~\ref{fgr:nex}) were already identified as not being major contributors to the bipolar doping phenomenon. The results of these calculations, shown in Fig. \ref{fgr:dep}, indicate that V\textsubscript{Cu} formation energy is sharply lowered towards the surface, whereas the Cu\lo{i} formation energy is only slightly lowered, and even increased for a Cu adatom on top of the surface. This finding implies that the prevalence of Cu\lo{i} over V\lo{Cu} in the bulk (Fig.~\ref{fgr:enth}) could be inverted close to the growing surface, creating a net p-type doping. It would then depend on kinetic factors such as diffusivity, temperature, and growth rates whether the defect concentrations at the surface would be dominated by Cu\lo{i}defects (i.e. excess Cu adatoms) or by V\lo{Cu} defects (i.e. excess nitrogen adatoms that do not recombine and desorb). To explore this hypothesis, we took a kinetic approach to describing the surface concentration of Cu\lo{i} and V\lo{Cu} defects as a function of temperature during growth. 

\begin{figure*}[t]
\centerline{\includegraphics[width=14cm]{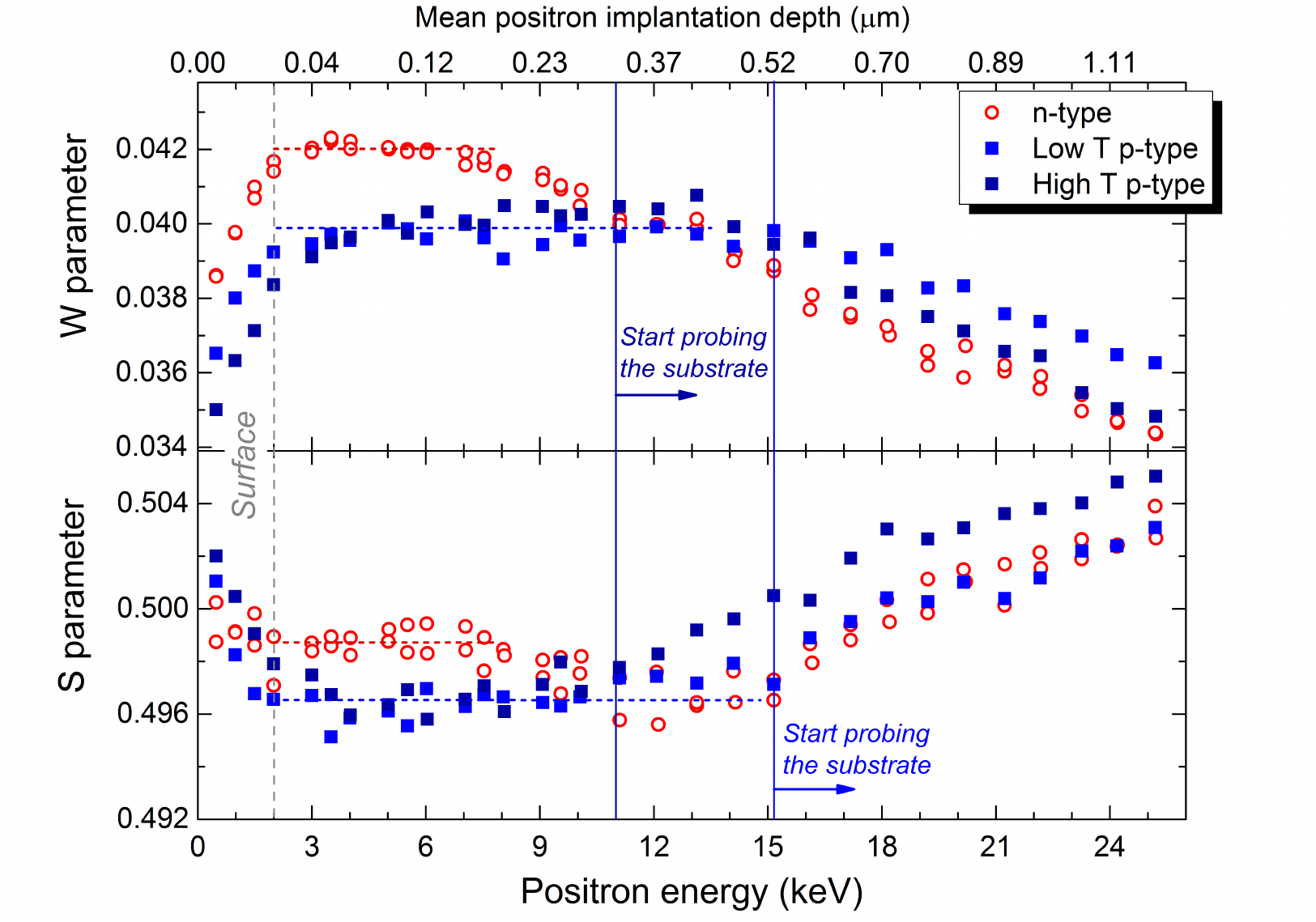}}
\caption{Energy-dependent positron annihilation experiments with p-type and n-type Cu\lo{3}N samples. As the positron penetration depth increases, the n-type sample's response becomes more similar to the p-type samples' response, before both of them trends towards the substrate. The dashed lines represents the plateaus defined in the text.}
  \label{fgr:pos}
\end{figure*}

\subsection{Defect Kinetics}

From the discussion above it is evident that the observed p-type character of Cu\lo{3}N at elevated deposition temperature, and the switch to the n-type character at low temperature (Fig.~\ref{fgr:trans}), both cannot be explained by either bulk (Fig.~\ref{fgr:enth}) or surface (Fig.~\ref{fgr:dep}) defect thermodynamics. Hence, we invoke here a surface kinetic model to explain the experimentally observed Cu\lo{3}N bipolar doping behavior. Before describing the proposed surface kinetic model, it is important to note that Cu\lo{3}N would not exist in the first place in the absence of relatively high bulk kinetic barriers, since it is a thermodynamically metastable material with +0.8 eV/f.u heat of formation.\cite{caskey2014} Thus, it is plausible that there are other lower-height kinetic barriers that control doping behavior below the bulk decomposition temperature, such as surface diffusion barriers. This is particularly true in this case of Cu\lo{3}N thin films for which the surface to volume ratio is larger than in the bulk.

First, we explain p-type conduction in Cu\lo{3}N at elevated substrate temperatures (50--120\degree C, Fig.~\ref{fgr:trans}a). Since this temperature range lies immediately below the decomposition temperature of bulk Cu\lo{3}N, diffusion in the bulk must be frozen, whereas surface diffusion may still occur. Hence, surface contributions ($\Delta H_{Surf}$) to the bulk defect formation enthalpies ($\Delta H_{Bulk}$) need to be taken into account to determine the resulting carrier concentration: $\Delta H_{Tot}=\Delta H_{Bulk}+\Delta H_{Surf}$, where $\Delta H_{Tot}$ is the resulting defect formation enthalpy that defines the carrier concentration.\cite{lany2007} Subtracting the calculated 1.5 eV surface contribution (Fig.~\ref{fgr:dep}) from the bulk formation energy of the V\lo{Cu} acceptor (Fig.~\ref{fgr:enth}), and adding the corresponding contribution for the Cu\lo{i} donor, makes the V\lo{Cu} acceptor the most energetically favorable defect under elevated growth temperature conditions. This enhanced V\lo{Cu} acceptor formation and suppressed Cu\lo{i} donor formation explains the measured p-type conductivity in Cu\lo{3}N in the 50--120\degree C substrate temperature range (Fig.~\ref{fgr:trans}a).

\begin{figure}[t]
\centerline{\includegraphics[width=9cm]{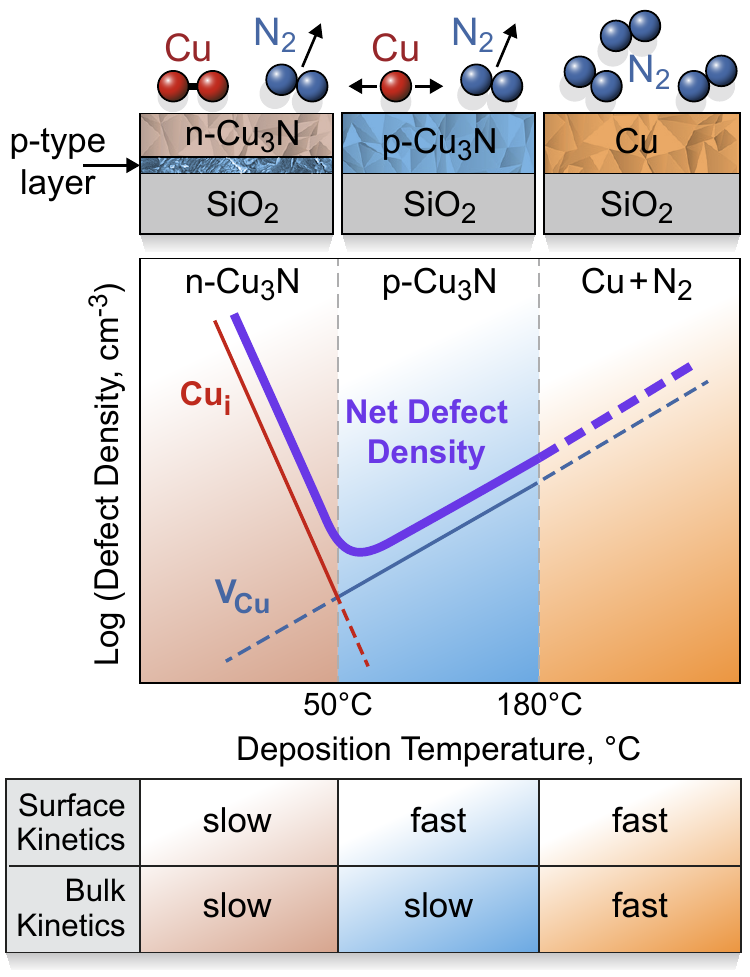}}
\caption{Schematic illustration of the current understanding of bipolar self-doping in Cu\lo{3}N.  This kinetics-driven doping mechanism explains the phenomenon as arising from a balance between surface and bulk kinetic processes. When surface diffusion is dominant and bulk diffusion is frozen (50--120\degree C), p-type Cu\lo{3}N is observed. When both surface and bulk diffusion is kinetically limited (no active heating), n-type Cu\lo{3}N is observed. See manuscript text for further explanation.}
  \label{fgr:kin}
\end{figure}

Next, we address n-type conductivity at lower substrate temperature ($\sim$35\degree C, Fig.~\ref{fgr:trans}a). It is likely that in this temperature regime, equilibration of the surface defects as discussed above does not fully occur. In this case, a non-equilibrium defect distribution becomes frozen in the bulk when the next monolayer of Cu and N atoms arrives. Note that Cu atoms are more likely to be frozen in the bulk than N atoms, because the Cu-Cu diatomic species probably has a longer residence time at the surface than the gaseous N-N species (i.e. N\lo{2} molecule). From a kinetic point of view, atom concentration (c) at the surface of the growing Cu\lo{3}N film at any given time is given by $c\approx c_0 e^{-Kt}$, where $K=K_0 e^{-Ea/k_bT}$ is the T-dependent desorption rate determined by activation energy $E_a$ (proportional to residence time of the atoms at the surface). This inherent Cu vs.~N asymmetry leads to enhanced formation of Cu\lo{i} donors, and thus produces n-type conductivity at these lower substrate temperatures. 

The kinetic character of the hypothesis proposed above to explain the temperature-dependence of Cu\lo{3}N conductivity type would imply that the deposition time (or rate) is another important variable in this kinetically-limited deposition growth process. Indeed, the rate dependence of the Cu\lo{3}N phase formation has been previously reported for films grown at elevated substrate temperatures.\cite{caskey2014} As for the lower-temperature phase-pure Cu\lo{3}N thin films reported here, the results of positron annihilation spectroscopy (PAS) suggest the presence of a p-type layer at the film/substrate interface of nominally n-type films. Fig.~\ref{fgr:pos} shows the W and S parameters plotted as a function of positron irradiation energy for n-type (red circles) and p-type (blue squares) Cu\lo{3}N films, respectively. These results indicate that the positron annihilation characteristics are homogeneous throughout the p-type samples' depth, while a defect inhomogeneity profile is observed through the thickness of the n-type samples. The first 150 nm of the n-type sample at the interface with the substrate are different than the rest of the 500 nm thick film. The S and W values form a plateau (shown with a dashed line) at the top of the film, then move toward the S and W values of the p-type films when proceeding towards the film/substrate interface. Hence, the defect structure of the nominally n-type sample resembles that of the p-type samples at the interface with the substrate. The presence of this interfacial p-type layer can be attributed to longer time spent at the growth temperature than the rest of the n-type film, supporting the hypothesis that the n-type doping is enabled by sluggish kinetics at this lower deposition temperature. 

It should be noted that detailed positron annihilation-based identification of the vacancy-type defects in the films in this work was not possible due to the lack of a reference sample with low defect density. For an additional representation of the PAS data, showing the variation of W with S for n-type and p-type samples, please see Fig.~S1 of the Supplemental Information.

Our current understanding of bipolar self-doping in Cu\lo{3}N is summarized in Fig.~\ref{fgr:kin}. At lower substrate temperatures, Cu\lo{3}N shows n-type conductivity due to Cu\lo{i} defects frozen in the bulk as a result of surface diffusion barriers. As the growth temperature increases, such that the thermal energy becomes comparable to the surface diffusion barriers, the conductivity switches from n-type to p-type due to the decreased formation enthalpy of V\lo{Cu}, and increased formation enthalpy of Cu\lo{i}, at the growing Cu\lo{3}N surface (Fig.~\ref{fgr:dep}). Finally, at the highest substrate temperature, Cu\lo{3}N decomposes into metallic Cu and N\lo{2},\cite{caskey2014} which corresponds to bulk equilibration of this metastable semiconducting material. Note that this mechanistic understanding is consistent with a number of experimental observations, including temperature-dependent conductivity type (Fig.~\ref{fgr:trans}), synchrotron-based NEXAFS measurements (Fig.~\ref{fgr:nex}), depth-dependent positron annihilation (Fig.~\ref{fgr:pos}), and Cu\lo{3}N growth/decomposition experiments (Ref. 7).\nocite{caskey2014}

\section{Conclusions}

We have presented an explanation for bipolar self-doping in the metastable material Cu\lo{3}N. Using a combination of high-throughput synthesis, advanced characterization, and different computational approaches, we were able to address the outstanding question of why bipolar self-doping behavior is observed in this material. The proposed mechanism implicates two native point defects, specifically the donor defect Cu\lo{i} and the acceptor defect V\lo{Cu}, as giving rise to n-type and p-type Cu\lo{3}N, respectively. Doping type was found to be temperature-dependent, with p-type doping observed for samples grown at elevated temperature (50--120\degree C) and n-type doping observed for samples grown with no active heating ($\sim$35\degree C). On average, p-type films in this work exhibited carrier density on the order of 10\hi{15}--10\hi{16} holes/cm\hi{3} with $\mu=$ 0.1--1.5 cm\hi{2} V\hi{-1} s\hi{-1} and n-type films exhibited $\sim$10\hi{17} e\hi{-}/cm\hi{3} with $\mu=$ 1.5 cm\hi{2} V\hi{-1} s\hi{-1}. Additionally, both the donor and the acceptor defects (Cu\lo{i} and V\lo{Cu}) were found to be present in both conduction types of Cu\lo{3}N based on NEXAFS analysis, indicating that competing defect formation as a function of temperature determines the ultimate majority carrier type. This hypothesis was supported by comparing bulk and surface defect formation enthalpies, in which it was shown that the acceptor defect formation is preferentially enhanced and the donor defect formation is preferentially suppressed at elevated growth temperatures just below the decomposition temperature of Cu\lo{3}N. The kinetics-driven doping mechanism presented herein explains this competition as a balance between surface and bulk kinetic processes. When surface diffusion is dominant and bulk diffusion is frozen (50--120\degree C), p-type Cu\lo{3}N is observed. When both surface and bulk diffusion is kinetically limited (no active heating), n-type Cu\lo{3}N is observed. This mechanistic understanding of copper nitride bipolar doping is consistent with several experimental observations, both from this work and previous works.\cite{caskey2014,zakutayev2014} Overall, the results of this work highlight the importance of kinetic processes in the defect physics of metastable materials, and provide a framework that can be applied when considering the properties of such materials in general.

\section{Acknowledgements} 

This work was supported by the U.S. Department of Energy under Contract No.~DE-AC36-08GO28308 with the National Renewable Energy Laboratory (NREL). A.N.F. was supported by the Renewable Energy Materials Research Science and Engineering Center under Contract No.~DMR-0820518 at the Colorado School of Mines. Computational resources utilized in this work were sponsored by the Department of Energy's Office of Energy Efficiency and Renewable Energy, loacted at NREL. Use of the Stanford Synchrotron Radiation Lightsource at SLAC National Accelerator Laboratory, is supported by the U.S. Department of Energy, Office of Science, Office of Basic Energy Sciences under Contract No.~DE-AC02-76SF00515.The authors would like to thank Celeste Melamed for providing the combinatorial grid diagram in Fig.~\ref{fgr:combi}. Thanks also to Alfred Hicks at NREL for the summary kinetic diagram in Fig.~\ref{fgr:kin}. Finally, the authors would like to thank Dr. Julien Vidal at EDF R\&D in Chatou, France for many helpful discussions.

\balance 
\bibliographystyle{rsc} 
\bibliography{Cu3N} 

\end{document}